\documentclass[conference]{IEEEtran}
\usepackage{cite}
\usepackage{amsmath,amssymb,amsfonts}
\usepackage{algorithmic}
\usepackage{graphicx}
\usepackage{textcomp}
\usepackage{xcolor}
\def\BibTeX{{\rm B\kern-.05em{\sc i\kern-.025em b}\kern-.08em
T\kern-.1667em\lower.7ex\hbox{E}\kern-.125emX}}

\usepackage{url}
\usepackage[]{fancyhdr} % 
\newcommand{\changefont}{\fontsize{9}{9}\selectfont}
\usepackage{nomencl}
\makenomenclature
\usepackage{etoolbox}
\renewcommand\nomgroup[1]{%
    \vspace{2ex}%
    \item[%
    \ifstrequal{#1}{A}{\emph{Sets and Indices}}{%
    \ifstrequal{#1}{B}{\emph{Parameters}}{%
    \ifstrequal{#1}{C}{\emph{Variables}}{%
    \ifstrequal{#1}{D}{\emph{Functions}}{}}}}%
    ]%
}
\usepackage{mathptmx}
\usepackage[short, nocomma]{optidef}
\usepackage{algorithm}
\usepackage{algorithmic}
\usepackage[hidelinks, breaklinks]{hyperref}
\begin{document}

%%%%% TITLE
\title{A Privacy-Preserving Energy Management System
for Cooperative Multi-Microgrid Networks\\}

%%%%% AUTHORS
\author{\IEEEauthorblockN{Carlos Ceja-Espinosa, Mehrdad Pirnia, %
Claudio A. Cañizares}
\IEEEauthorblockA{\textit{Department of Electrical and Computer Engineering} \\
\textit{University of Waterloo}\\
Waterloo, Ontario, Canada \\
ccejaesp@uwaterloo.ca, mpirnia@uwaterloo.ca, ccanizares@uwaterloo.ca}
% \and
% \IEEEauthorblockN{Author 2}
% \IEEEauthorblockA{\textit{Electrical and Computer Engineering} \\
% \textit{University of Waterloo}\\
% Waterloo, Canada \\
% @uwaterloo.ca}ñ
% \and
% \IEEEauthorblockN{Author 3}
% \IEEEauthorblockA{\textit{Electrical and Computer Engineering} \\
% \textit{University of Waterloo}\\
% Waterloo, Canada \\
% @uwaterloo.ca}
}

\maketitle
%%% FROM CONFERENCE TEMPLATE
\thispagestyle{fancy}
\pagestyle{fancy}
%%%

%%%%% ABSTRACT
\begin{abstract}
This paper presents an Energy Management System (EMS) that considers power
exchanges between a set of interconnected microgrids (MGs) and the main grid,
in the context of Multi-MG (MMG) systems.
The model is first formulated as a centralized optimization problem, which is
then decomposed into subproblems corresponding to each MG, using Lagrangian
relaxation, and solved through a distributed approach using a
subgradient method.
The proposed model determines the power exchanges minimizing the operation
cost of each MG, considering grid constraints and preserving the privacy of
each MG by not revealing their generation cost and demand information.
The distributed approach is validated with respect to the centralized problem,
and various case studies are presented to demonstrate the performance of the
proposed approach, comparing the costs of the MGs operating individually and
cooperatively. The results show that all MGs in the MMG system improve their
cost as consequence of the power exchanges, thus demonstrating the advantages
of interconnecting MGs.
\end{abstract}

%%%%% INDEX TERMS

%%%%% NOMENCLATURE
%%%%% SETS
\nomenclature[A]{$ M $}{Set of microgrids}

%%%%% INDICES
\nomenclature[A]{$ m, n $}{Index for each microgrid}

%%%%% PARAMETERS
\nomenclature[B]{$ a_{m} $}{Coefficient of the quadratic term of the cost
function of microgrid $m$ $ [\mathrm{\$/kW^2}] $}

\nomenclature[B]{$ b_{m} $}{Coefficient of the linear term of the cost
function of microgrid $m$ [\$/kW]}

\nomenclature[B]{$ c_{m} $}{Coefficient of the constant term of the cost
function of microgrid $m$ $ [\$] $}

\nomenclature[B]{$ \overline{P}_{m} $}{Maximum total power generation of
microgrid $m$ [kW]}

\nomenclature[B]{$ D_{m} $}{Power demand in microgrid $m$ [kW]}

\nomenclature[B]{$ \overline{P_{m}^{\textrm{PCC}}} $}{Maximum power transfer
capacity through the point of common coupling of microgrid $m$ [kW]}

\nomenclature[B]{$ c_{d,m} $}{Cost of buying electricity from the main grid for
microgrid $m$ [\$/kW]}

\nomenclature[B]{$ c_{m,d} $}{Cost of buying electricity from microgrid $m$ for
the main grid [\$/kW]}

\nomenclature[B]{$ \alpha $}{Update step size for the subgradient method}

\nomenclature[B]{$ \beta $}{Scaling factor of the transfer cost function}

%%%%% VARIABLES
\nomenclature[C]{$ P_{m} $}{Power generation in microgrid $m$ [kW]}

\nomenclature[C]{$ P_{m,n}^{e} $}{Power exchange between microgrid $m$ and
microgrid $n$ [kW]}

\nomenclature[C]{$ P_{d,m} $}{Power sold by the main grid to microgrid $m$ [kW]}

\nomenclature[C]{$ P_{m,d} $}{Power bought by the main grid from microgrid $m$
[kW]}

\nomenclature[C]{$ \lambda_{m} $}{Selling price of microgrid $m$ to other
microgrids [\$/kW]}

\nomenclature[C]{$ \varepsilon_{m} $}{Total power sold by microgrid $m$ [kW]}

%%%%% FUNCTIONS
\nomenclature[D]{$ C_{m} $}{Generation cost function of microgrid $m$ [\$]}

\nomenclature[D]{$ \gamma $}{Power transfer cost function [\$]}
\printnomenclature

%%%%% CONTENT
\section{Introduction}

In recent years, environmental concerns have motivated a gradual transformation
of power systems, shifting from fossil fuel-based energy sources, such as coal
and natural gas, to Renewable Energy Sources (RES), such as wind and solar
energy.
In this context, microgrids (MGs) have been proposed as a solution to integrate
these RES into existing power systems in a reliable and efficient manner.
An MG can be described as a ``cluster of loads, distributed generation (DG)
units and energy storage systems (ESSs) operated in coordination to reliably
supply electricity, connected to the host power system at the distribution
level at a single point of connection, the point of common coupling (PCC)''
\cite{Olivares2014}. MGs should be capable of operating both in isolated
and grid-connected modes, thus enhancing the system resilience and reducing
power transmission losses.

Connected MGs to form a multi-MG (MMG) system, working in coordination to
further improve the operation of power systems, have recently attracted
significant attention, with ongoing research targeting different aspects of
the problem.
A comprehensive discussion on the potential of networked MGs to support
the resiliency of power systems in the case of extreme events is presented
in \cite{Li2017}. The resiliency is enhanced by connecting geographically
close MGs to form an MMG system, and coordinating power transfers to exploit
the heterogeneity of each MG and ride through extreme events with the
support of all MGs. A detailed survey focusing on distributed
control and communication strategies in networked MGs is presented in
\cite{Zhou2020}, which reiterates the benefits of networked MGs and describes
characteristics and issues with the communication network in MMG systems.
Multi-agent system (MAS)-based distributed coordination
control strategies are reviewed in \cite{Han2018}, and topology and
mathematical models and techniques are described, such as graph theory,
non-cooperative game models, genetic algorithms, and particle swarm
optimization; furthermore, distributed consensus protocols and techniques to
mitigate communication delays are discussed. In \cite{Alam2019}, a discussion
on the architecture, control, and communication techniques proposed in the
literature for networked MGs is presented, where benefits and
challenges in the implementation of such systems are outlined. An overview
of proposed control solutions is provided in \cite{Vasilakis2020}, where
popular techniques such as model predictive control (MPC) and reinforcement
learning (RL) are explained in detail.
These references constitute a comprehensive review of the current state of
research and development in MMG systems, and clearly establish
their key features of interest to evolving power systems, i.e., enabling the
safe and reliable inclusion of renewable energy sources with high penetration
levels, enhancing the resilience of power systems, and
exploiting the heterogeneity of local generation and demand to attain technical
and economic benefits for the overall operation of the system.

In general, the existing literature falls into the categories of system
planning, voltage and frequency control, communication in power sharing,
service restoration, stability enhancement, and optimal energy management.
The present work is in the last category, where most of the literature
deals with mechanisms of coordination for operating the MMG system optimally,
the modeling of the energy management system (EMS) as centralized or
distributed optimization problems, the solution technique of such problems,
and the characteristics of different system configurations \cite{Zou2019}.
In this context, in \cite{Xu2020}, a distributed and robust EMS for networked
hybrid ac/dc MGs is proposed, coordinating the energy sharing through the
dc network, to minimize the power transmission losses. A distributed solution to
preserve the privacy of each MG is implemented using the Alternating-Direction
Method of Multipliers (ADMM) algorithm, which is applicable to this model
because all the energy exchanges occur through the dc network, i.e., MGs only
need to update their energy exchange schedule with the network, and not other
MGs.

A two-layer system to enable peer-to-peer electricity sharing is proposed in
\cite{Luo2019}, where the first layer corresponds to a multi-agent coalition
mechanism that allows ``prosumers'' to negotiate electricity trading;
the second layer consists of a blockchain-based mechanism that ensures the
secure settlement of the transactions established in the first layer.
A peer-to-peer trading mechanism that uses locational marginal prices to
compute network usage charges is presented in \cite{Kim2020}, with a
decentralized configuration that focuses on preferences of each individual
peer. The proposed mechanism is an iterative price-adjusting process that 
matches the seller and buyer prices for all energy trades.
In \cite{Hussain2018}, a ``nested'' day-ahead scheduling model for networked
MGs is proposed, in which MGs are nested based on load priority.
The layered structure enhances the system resiliency, and the operation costs
are comparable to a centralized approach. However, this nested configuration
significantly increases the complexity of the problem, which might deter its
use in practical systems.

A decentralized energy management framework formulated as a bi-level quadratic
optimization problem is developed in \cite{Liu2020}. The upper level
corresponds to the distribution system, and the lower level to each individual
MG; the two levels are linked through the clearing price determined in
the upper level optimization. A distributionally robust optimization algorithm
is developed, which performs better than a stochastic programming
approach, with results showing that the operation cost of the distribution
system is greatly reduced when compared with fixed pricing schemes.
The benefits of cooperative operation of MGs are explored in
\cite{Parisio2017}, where an MPC approach to control the operation of MGs
is proposed. A three-stage distributed algorithm is developed, starting with
local optimizations for each MG, followed by a power distribution stage that
coordinates all MGs according to the energy plan of an aggregator, and a
final stage to redistribute power deviations from the aggregator's profile.
Numerical experiments show that through cooperation, MGs reduce their
operation cost and the amount of energy exchanged with the grid, when compared
to a centralized solution. A similar technique is presented in \cite{Du2019},
where a distributed MPC dispatch process is implemented, with upper and lower
layers for the energy exchanges and supply and demand balance, respectively.
Results show once again that using the coordinated MPC approach reduces the
daily operational costs.

In \cite{Wang2015}, a coordination strategy for networked MGs
formulated as a stochastic bi-level problem is proposed, where the first stage
determines base generation setpoints according to the load and forecasts of
non-dispatchable units, while the second stage adjusts the generation output
based on realized scenarios. Simulations show that the stochastic model can
lead to higher profit for entities with renewable generation, as it considers
the corresponding uncertainties more accurately. However, this is not a
distributed solution to the energy management problem.
A decentralized framework based on ADMM to coordinate power exchanges between
the distribution system and MMGs is presented in \cite{Gao2018}, where
each entity is modeled as a two-stage robust optimization problem to account
for uncertainties in renewable generation and demand,
assuming that only one MG is connected to the distribution system.
For this reason, the information exchange just involves the distribution
system and a single MG, which makes it possible to apply the ADMM
algorithm. Similar approaches, with different models and modified versions of
ADMM algorithms are presented in \cite{Wang2016} and \cite{Waqar2021}.

A distributed optimization framework for energy trading between MGs is
presented in \cite{Gregoratti2015}, which considers a simple MMG model,
and implements a subgradient-based algorithm that requires minimal
information exchange, thus preserving the privacy of each MG.
In \cite{Wang2018}, a mechanism based on Nash bargaining theory to incentivize
energy trading among MMGs is presented. The bargaining problem is decomposed
into two sequential problems, one for social cost minimization
and one for trading benefit sharing, and a decentralized solution using an ADMM
algorithm is implemented. Results indicate that the overall costs of the
MGs participating in energy trading are reduced.

From the reviewed literature, it is clear that the topic of networked MGs
is relevant for the future development of smart grids. However,
most studies formulate a particular model suitable for the proposed
solution approach, with results reported on diverse test systems, making the
comparison between techniques very difficult.
Hence, the contributions and objectives of this paper are as follows:
\begin{itemize}
\item Develop a centralized, decomposable, and more realistic model for the
optimal and coordinated operation of MMGs, exchanging power among each other
and with the main grid, while considering their physical interconnections and
their limits, which most existing papers on this topic ignore, even though
this is the case in practice.
\item Implement a distributed model for optimal
scheduling of the participating MGs, while preserving their privacy,
and coordinating their cooperative operation.
\item Test and validate the proposed model in an MMG system, considering its
physical grid layer and associated constraints, while demonstrating the
advantages of the MGs interconnection.
\end{itemize}

The rest of the paper is organized as follows: Section~II provides a brief
overview of MMG systems and distributed optimization techniques.
Section~III presents the proposed centralized model for the MMG system
and the distributed approach, including the required modification to reach
convergence and the decomposition procedure. In Section~IV, case studies are
presented and discussed, testing and validating the proposed model. Finally,
Section~V summarizes the conclusions and contributions of this paper,
and identifies future research objectives.

\section{Background}

\begin{figure}[tb]
\centering
\includegraphics[scale=1.0]{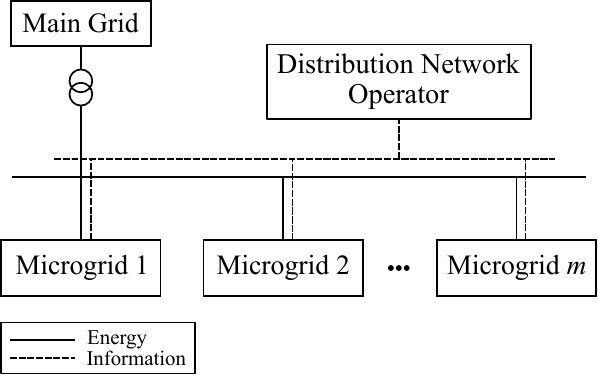}
\vspace{0ex}
\caption{Typical MMG system in practice.}
\label{Fig: Multi_Microgrid_System}
\end{figure}

An MMG system can be described as a network formed by multiple individual MGs
that are geographically close and connnected to the same distribution bus,
with the ability to exchange power among MGs and/or the distribution grid at
the PCC \cite{Alam2019}.
MMG systems offer several advantages, such as more economic operation in
grid-connected mode, mitigation of congestion in distribution lines,
and the capability to isolate from compromised sections of the grid when faults
or natural disasters occur \cite{Zou2019}.
All MGs in an MMG system should be able to exchange energy with the main
power grid, under the supervision of the distribution system operator,
assuming three common topologies of MMGs: radial topology, in which
each MG is connected directly to the main grid, with energy exchanges taking
place through the distribution bus; daisy chain topology, in which energy
and information is exchanged bidirectionally between adjacent MGs; and mesh
topology, in which all MGs are interconnected to each other and the main grid
directly, to exchange energy and information. Of these topologies, the
radial topology shown in Fig.~\ref{Fig: Multi_Microgrid_System} is the most
realistic, considering the physical connection of MGs to the distribution
system.

The operation of an MMG system with a mesh topology is commonly studied
in the literature, for example in \cite{Ouammi2015} and \cite{Rahbar2018},
which provide a valuable mathematical insight into the MMG EMS problem.
While this configuration is more general in theory, and supports the
convergence of the distributed optimization algorithm in a more straightforward
fashion \cite{Gregoratti2015}, such topologies are not common in practice.
Therefore, the radial topology is implemented in our proposed distributed
approach to enhance the practicality of our model.

Recent developments in communication technologies have resulted in
emergence of networked systems, in which interconnected subsystems cooperate
to achieve a global objective beneficial to all participants.
The concept of MMG networks is a good example of such a system, with an
inherent decentralized structure that can be advantageously exploited.
Several solution methods and algorithms have been reported in the
literature, based on distributed optimization techniques \cite{Yang2019},
suitable for these types of problems.

In this paper, a procedure known as dual decomposition is used to solve
the MMG operation model. This procedure has been widely applied
to various optimization problems that benefit from a distributed solution.
A generic description of dual decomposition, as presented in
\cite{Kazempour2019}, can be explained based on the following mathematical
model:
\begin{mini!}[2]
{x} {\sum_{i=1}^{I} f_{i}(x_{i})} {\label{Eqn: Primal_Objective}} {}
\addConstraint{g_{i}(x_{i})}{\leq A_{i} \quad \forall \ i
\label{Eqn: Primal_Constraints}}{}
\addConstraint{\sum_{i=1}^{I} h_{i}(x_{i})}{\leq B
               \label{Eqn: Primal_Constraints_Complicating}\quad}{}
\end{mini!}
where the objective function $f_{i} (x_{i})$ is a convex, quadratic, and
separable function with respect to variables $x_{i}$.
Constraints \eqref{Eqn: Primal_Constraints} are also separable (one constraint
for each $i$), but constraints \eqref{Eqn: Primal_Constraints_Complicating}
are complicating constraints because they are not separable and prevent
the decomposition of the problem.
In order to relax the complicating constraints, a dual decomposition procedure
is used by dualizing~\eqref{Eqn: Primal_Constraints_Complicating},
as follows \cite{Boyd2010}:
\begin{mini!}
{x_{i}} {\sum_{i=1}^{I} f_{i}(x_{i})
+ \lambda \bigg[ B - \sum_{i=1}^{I} h_{i}(x_{i}) \bigg]}
{\label{Eqn: LR_Objective}}
{\underset{\lambda}{\textrm{max}}\bigg\{}
\addConstraint{g_{i}(x_{i})}
{\leq A_{i}\quad \label{Eqn: LR_Constraints}}
{\forall i \bigg\}}
\end{mini!}
This problem is not decomposable, due to the Lagrange multiplier (or dual
variable) $\lambda$ in the objective function. To solve this issue, the problem
is relaxed by fixing $\lambda$ to a constant value $\overline{\lambda}$, i.e.,
turning it into a parameter, with the relaxed problem being decomposable into
the following $i$ subproblems:
\begin{mini!}
{x_{i}}{f_{i}(x_{i}) - \overline{\lambda} h_{i}(x_{i})}
{\label{Eqn: LR_Objective_Rel}}{}
\addConstraint{g_{i}(x_{i})}{\leq A_{i}
\quad \label{Eqn: LR_Constraints_Rel}}{}
\end{mini!}

In dual decomposition, the maximization over $\lambda$ is carried out through
the iterative solution of the obtained subproblems, updating the fixed dual
variable in each iteration with an appropriate technique, such as the
subgradient or cutting plane methods \cite{Conejo2006}.
The subgradient method is the most straightforward approach, based on the
following update step:
\begin{equation}
    \lambda^{k+1} = \lambda^{k} + \alpha^{k} s^{k}
    \label{Eqn: Subgradient_Update}
\end{equation}
where $\alpha$ is the step size, $s$ is the subgradient of
\eqref{Eqn: LR_Objective_Rel}, and $k$ is the iteration counter. The step size
can be chosen according to different rules, as described in \cite{Boyd2003}.
For a constant step size, the subgradient algorithm is guaranteed to converge
within some range of the optimal value in a finite number of steps,
depending on the step size.
The iterative procedure starts by initializing the dual variable at some
fixed value, solving each subproblem independently with the dual variable as
parameter, and updating the dual variable with the obtained solutions using
\eqref{Eqn: Subgradient_Update}. The process is then repeated by solving the
subproblems with the updated dual variables, until convergence is reached.
This distributed solution approach is used here.

\section{Methodology}

The MMG system considered in this work is comprised of $m$ MGs connected to
the main grid, as shown in Fig.~\ref{Fig: Multi_Microgrid_System}. It is
assumed that each MG can buy or sell power from and to the main grid, as well
as exchange power with all other MGs through the main grid, but not directly,
as per the most realistic MMG topology, which is not the case in most
publications on this topic. In this system, it is important to consider the
capacity limits of the connection of each MG; furthermore, information exchange
is necessary to coordinate the power exchanges among MGs.

Two approaches, namely, centralized and distributed, can be used to solve the
EMS problem within MMGs. This paper focuses on the formulation of an MMG EMS
which is suitable for decomposition and the subsequent application of a
distributed algorithm, with the goal of finding optimal or near-optimal
solutions that preserve the privacy of each MG. Hence, the distribution and
communication networks, and other components in an MMG system,
such as power electronics that interface each MG with the main grid,
are not directly considered.
Furthermore, a simplified model is adopted, in which the
only components within each MG are a single thermal generation unit and
a given demand. Additional components of an MG, such as renewable energy
generators and energy storage devices, can be readily included in the
formulation, but are not considered here since the paper focuses on developing
an appropriate decomposition approach suitable for MMG system operation that
considers the physical grid interconnection and limitations, unlike other
papers in the literature, particularly \cite{Gregoratti2015}, which is the
basis for the proposed decomposition technique.
For the same reason, the model is developed for a single scheduling period
(assumed to be one hour), but can be readily extended to consider several
time periods. It is important to note that the proposed MMG EMS decomposition
procedure and distributed algorithm are applicable even if the aforementioned
modifications are introduced in the model.

\subsection{Centralized Problem}

Assuming that there is only one thermal generator such as a diesel unit in
each MG, a single generation cost function $C_{m}$ for the entire MG can be
considered:
\begin{equation}
C_{m} (P_{m}) = a_{m} P_{m}^{2} + b_{m} P_{m} + c_{m}
\quad \forall \ m
\label{Eqn: Generation_Cost}
\end{equation}
where $a_{m}$, $b_{m}$, and $c_{m}$ are the quadratic, linear, and constant
coefficients of the cost function, and $P_{m}$ is the power output of the
generator, as defined in the Nomenclature section for this and the next
equations. Then, the cost minimization (primal problem) for the operation of a
system comprised of $m$ MGs can be formulated as follows:
\begin{mini!}[2]
{\substack{P_{m}, P_{d,m}, \\ P_{m,d}, P_{m,n}^{e}}}
{\sum_{m} [C_{m} (P_{m}) + c_{d,m} P_{d,m} - c_{m,d} P_{m,d}]
\label{Eqn: Centralized_Objective}}
{\label{Eqn: Centralized_Problem}} {}
\addConstraint{P_{m} + \sum_{n} P_{n,m}^{e} + P_{d,m} = D_{m}
+ \sum_{n} P_{m,n}^{e}}{}{\nonumber}
\addConstraint{\hspace{10.15em} + P_{m,d} \quad \forall \ m}
{\label{Eqn: Power_Balance}}
{}
\addConstraint{0 \leq P_{m} \leq \overline{P}_{m} \quad \forall \ m}
{\label{Eqn: Generation_Limits}}
{}
\addConstraint{\sum_{n} P_{m,n}^{e} + P_{m,d}
\leq \overline{P_{m}^{\textrm{PCC}}} \quad \forall \ m}
{\label{Eqn: PCC_Upper_Limit}}
{}
\addConstraint{\sum_{n} P_{n,m}^{e} + P_{d,m}
\leq \overline{P_{m}^{\textrm{PCC}}} \quad \forall \ m}
{\label{Eqn: PCC_Lower_Limit}}
{}
\addConstraint{P_{m,n}^{e} P_{n,m}^{e} = 0 \quad \forall \ m, n}
{\label{Eqn: Pe_Complementarity}}
{}
\addConstraint{P_{m,n}^{e}, \ P_{d,m}, \ P_{m,d} \geq 0 \quad \forall \ m, n}
{\label{Eqn: Vars_Nonnegativity}}
{}
\end{mini!}
The selling and buying prices of the main grid to and from MG $m$ are denoted
by $c_{d,m}$ and $c_{m,d}$, respectively. Variables $P_{d,m}$ and $P_{m,d}$
represent the power sold and bought by the main grid to and from MG $m$,
respectively, and $P_{m,n}^{e}$ correspond to the variable indicating power
sold by MG $m$ to MG $n$. Similarly, variable $P_{n,m}^{e}$ corresponds
to the power sold by MG $n$ to MG $m$ in the set of MGs $M$, assuming that
$m \neq n$. $D_{m}$ is the demand in MG $m$, and $\overline{P}_{m}$ and
$\overline{P_{m}^{\textrm{PCC}}}$ represent the maximum power generation and
power transfer capacity through the PCC of MG~$m$.

The objective function \eqref{Eqn: Centralized_Objective} considers the total
generation cost, cost of buying from the main grid, and profit from selling to
the main grid for all MGs.
The supply and demand balance is defined by \eqref{Eqn: Power_Balance} for each
MG, considering exchanges between MGs and imports/exports from
and to the main grid, with dual variables $\lambda_{m}$, which represent the
selling prices of electricity for each MG.
The generation limits of each MG are set by \eqref{Eqn: Generation_Limits},
and the upper and lower limits of power transfer capacity through the PCC of
each MG are enforced by \eqref{Eqn: PCC_Upper_Limit} and
\eqref{Eqn: PCC_Lower_Limit}. The simultaneous power exchanges in opposite
directions between MGs are prevented by \eqref{Eqn: Pe_Complementarity},
and all power exchange variables are defined as non-negative variables by
\eqref{Eqn: Vars_Nonnegativity}.
This centralized optimization problem requires a central entity
to have access to all the information of each MG (cost functions and demands,
in this case). Such setup is in general not feasible if each MG is owned by
independent entities, as expected in practice; hence, there is a need for a
distributed solution method to avoid the need for exchange of sensitive
information through a central operator.

\subsection{Distributed Approach}

To solve problem \eqref{Eqn: Centralized_Problem} in a distributed manner,
with minimum information exchange through a central entity, a decomposition
approach is used to separate the complete problem into subproblems that can
be solved locally by each MG. While the decomposition method in this paper is
based on the model and procedure proposed in \cite{Gregoratti2015}, the
model is modified here to reflect the realistic operation of MMGs by
including PCC capacity constraints and avoiding simultaneous transactions in
opposite directions among MGs.

According to \cite{Boyd2010}, the objective function of the problem must be
convex with respect to the coupling variables, which are the power exchange
variables. However, \eqref{Eqn: Centralized_Objective} is not convex with
respect to the $P^{e}$ variables, causing convergence issues when solving the
dual problem with the iterative procedure described in Section~II.
Therefore, a power transfer cost function is added in the objective function to
resolve this issue. This new term, which is assumed quadratic to convexify it,
represents the cost of power transfer for the use of the distribution network,
which should be paid to the distribution network operator, and is defined
as follows:
\begin{equation}
\gamma (P_{n,m}^{e}) = \beta (P_{n,m}^{e})^{2}
\label{Eqn: Transfer_Cost_Function}
\end{equation}
where $\beta$ is a scalar chosen as small as possible to reduce the impact of
this term in the objective function value, considering that the cost of power
transfer is low. The primal problem after adding this transfer cost to
the objective function then becomes:
\begin{mini!}[2]
{\substack{P_{m}, P_{d,m}, \\ P_{m,d}, P_{m,n}^{e}}}
{\nonumber \sum_{m} \big[ C_{m} (P_{m}) + \sum_{n} \gamma(P_{n,m}^{e})
\label{Eqn: Centralized_Objective_Nom}}
{\label{Eqn: Mod_Centralized}} {}
\breakObjective{+ c_{d,m} P_{d,m} - c_{m,d} P_{m,d} \big]}
\addConstraint{\eqref{Eqn: Power_Balance}\textrm{--}
\eqref{Eqn: PCC_Lower_Limit}, \ \eqref{Eqn: Vars_Nonnegativity}}
{\label{Eqn: Centralized_Constraints_Nom}} {}
\end{mini!}
Notice that the bilinear constraint \eqref{Eqn: Pe_Complementarity}
is no longer needed in the revised formulation because the added power transfer
cost function prevents simultaneous power exchanges in opposite directions
between MGs. This important observation allows decomposing the
problem by MGs, as \eqref{Eqn: Pe_Complementarity} is a complicating constraint
that prevents decomposing the problem.

The coupling variables $P_{m,n}^{e}$ in \eqref{Eqn: Power_Balance}
and \eqref{Eqn: PCC_Lower_Limit} do not allow the problem to be separable.
Thus, to decompose the problem, an auxiliary variable $\varepsilon_{m}$ is
introduced, to represent the total power sold by MG $m$ to all other MGs,
defined as follows:
\begin{equation}
\varepsilon_{m} = \sum_{n} P_{m,n}^e
\quad \forall \ m
\label{Eqn: Aux_Variable_Def}
\end{equation}
Then, problem \eqref{Eqn: Mod_Centralized} can be modified by adding
\eqref{Eqn: Aux_Variable_Def} as a constraint and changing
\eqref{Eqn: Power_Balance} and \eqref{Eqn: PCC_Upper_Limit}  to the following
constraints:
\begin{gather}
P_{m} + \sum_{n } P_{n,m}^{e} + P_{d,m}
= D_{m} + \varepsilon_{m} + P_{m,d}
\quad \forall \ m
\label{Eqn: Power_Balance_Mod} \\
\varepsilon_{m} + P_{m,d}
\leq \overline{P_{m}^{\textrm{PCC}}}
\quad \forall \ m
\label{Eqn: PCC_Upper_Limit_Mod}
\end{gather}
With this modification, \eqref{Eqn: Aux_Variable_Def} is now the only
complicating constraint, and therefore a decomposition method such as the
dual decomposition method described in Section~II can be applied.
Thus, after dualizing \eqref{Eqn: Aux_Variable_Def}, the Lagrangian dual
problem can be expressed as follows:
\begin{mini!}[2]
{\substack{P_{m}, P_{d,m}, \\ P_{m,d}, P_{n,m}^{e}}}
{\nonumber \sum_{m} \bigg[ C_{m} (P_{m}) + \sum_{n} \gamma(P_{n,m}^{e})
                           + c_{d,m} P_{d,m}}
{\label{Eqn: Dual_Lagrangian}}
{\underset{\lambda_{m}}{\textrm{max}}\bigg\{}
\breakObjective{\hspace{2.7em} - c_{m,d} P_{m,d} + \lambda_{m}
\bigg( \sum_{n} P_{m,n}^{e} - \varepsilon_{m} \bigg) \bigg]}
\addConstraint{\eqref{Eqn: Generation_Limits}, \
\eqref{Eqn: PCC_Lower_Limit}, \
\eqref{Eqn: Vars_Nonnegativity}, \
\eqref{Eqn: Power_Balance_Mod}, \
\eqref{Eqn: PCC_Upper_Limit_Mod}}
\addConstraint{\varepsilon_{m} \geq 0 \quad \forall \ m}
{\label{Eqn: Aux_LR}}
{\bigg\}}
\end{mini!}
This minimization problem can now be separated into subproblems corresponding
to each MG, as follows:
\begin{mini!}[2]
{\substack{P_{m}, P_{d,m}, P_{m,d}, \\ P_{n,m}^{e}, \varepsilon_{m}}}
{\nonumber C_{m} (P_{m}) + \sum_{n} \gamma(P_{n,m}^{e})
+ c_{d,m} P_{d,m} \label{Eqn: Objective_Subproblem}}
{\label{Eqn: Subproblem}} {}
\breakObjective{- c_{m,d} P_{m,d} + \sum_{n} \lambda_{n} P_{n,m}^{e}
- \lambda_{m} \varepsilon_{m}}
\addConstraint{P_{m} + \sum_{n} P_{n,m}^{e} + P_{d,m}
= D_{m} + \varepsilon_{m}}
{}
{\nonumber}
\addConstraint{\hspace{10.15em} + P_{m,d}}
{\label{Eqn: Power_Balance_Subproblem}}
{\quad :\lambda_{m}}
\addConstraint{0 \leq P_{m} \leq \overline{P}_{m}}
{\label{Eqn: Generation_Limits_Subproblem}}
{}
\addConstraint{\varepsilon_{m} + P_{m,d}
\leq \overline{P_{m}^{\textrm{PCC}}}}
{\label{Eqn: PCC_Upper_Limit_Subproblem}}
{}
\addConstraint{\sum_{n} P_{n,m}^{e} + P_{d,m}
\leq \overline{P_{m}^{\textrm{PCC}}}}
{\label{Eqn: PCC_Lower_Limit_Subproblem}}
{}
\addConstraint{P_{n,m}^{e}, \ P_{d,m}, \ P_{m,d},
\ \varepsilon_{m} \geq 0 \quad \forall \ n}
{\label{Eqn: Vars_Nonnegativity_Subproblem}}
{}
\end{mini!}

Note that \eqref{Eqn: Subproblem} is now a local problem solved by each MG $m$
independently, with the dual variable $\lambda_{m}$ of its supply-demand
balance constraint representing the selling price of each MG, and variable
$P_{m,n}^{e}$ belonging to the subproblems of the other $n$ MGs.
For example, assuming there are three MGs in the system, the variables
associated with power exchanges among MGs in the subproblem of MG1 would be
$P_{2,1}^{e}$, $P_{3,1}^{e}$, and $\varepsilon_{1}$, while the variables
associated with power exchanges among MGs for MG2 would be $P_{1,2}^{e}$,
$P_{3,2}^{e}$, and $\varepsilon_{2}$; a similar pattern would be observed for
the third MG. Thus, by separating the variables this way, each MG is optimizing
its own problem in terms of the individual power bought from all other MGs,
and the total power sold to all other MGs. The coupling constraint, which
contains terms associated to all MGs, is now separated into terms appearing in
each subproblem.

To solve the dual problem defined in \eqref{Eqn: Dual_Lagrangian}, the
subgradient method \cite{Boyd2003} is used, which is implemented through the
iterative procedure described in Algorithm~\ref{Alg: Subgradient_Method}.
\begin{algorithm}[t]
\caption{Dual Problem Solution}
\begin{algorithmic}[1]
\STATE Initialize selling prices $\lambda_{m}$
\REPEAT
\STATE MGs share $\lambda_{m}$ with each other
\STATE MGs solve \eqref{Eqn: Subproblem} independently
\STATE MGs share $P_{n,m}^{e}$ at the current $\lambda_{m}$
\STATE MGs update their selling prices:
\begin{equation}
    \lambda_{m} \leftarrow
    \lambda_{m} + \alpha \big(\sum_{n} P_{m,n}^e - \varepsilon_{m} \big)
    \label{Eqn: Update_Rule}
\end{equation}
\UNTIL{Convergence criterion is satisfied}
\end{algorithmic}
\label{Alg: Subgradient_Method}
\end{algorithm}
The multiplier update step performed in \eqref{Eqn: Update_Rule} corresponds
to the basic subgradient method. This algorithm is very sensitive to the
chosen step size~$\alpha$. Other methods proposed in the literature to update
the multipliers, such as cutting plane or bundle methods, may converge faster
than the subgradient method, but are more cumbersome to implement.
Therefore, for the sake of clarity, the subgradient method was chosen in
this paper. The other issue with the proposed method in
Algorithm~\ref{Alg: Subgradient_Method}, is choosing the initial value of
$\lambda_{m}$, which may be determined from existing and/or expected exchange
prices.

\section{Case Studies}

The solution of the centralized problem \eqref{Eqn: Centralized_Problem} is
used as reference to validate the proposed distributed model
\eqref{Eqn: Subproblem} for an MMG test system comprised of three MGs.
All simulations were performed on a personal computer with a 1.6~GHz processor,
and implemented using the Pyomo optimization modeling language.
The IPOPT~3.11.1 solver was used for the nonlinear centralized
problem \cite{Waechter2005}, and the Gurobi~9.5 solver was used for the
distributed problem \cite{Gurobi}.

Two different cases were simulated, namely, a base case, corresponding to the
data presented in Table~\ref{Tab: Data_Base}, and a stressed case in which
larger power exchanges were enforced by making the generation of one MG very
expensive. The data in Table~\ref{Tab: Data_Base} was
collected and adapted from \cite{Xu2020} and \cite{Liu2020}.
The selling electricity price from the main grid to MGs was assumed
$c_{d,m} = 0.082 \ \mathrm{\$/kW}$, as per \cite{ElectricityRates2022}, and the
selling electricity price from MGs to the main grid was assumed
$c_{m,d} = 0.05 \ \mathrm{\$/kW}$, which is lower than $c_{d,m}$ to incentivize
the MG exchanges.
Note that the units of these parameters are given in \$/kW because a single
scheduling period of one hour was assumed throughout the formulation.

In order to have a fair comparison between the centralized and distributed
solutions, results of the original centralized problem defined in
\eqref{Eqn: Centralized_Problem}, the modified centralized problem including
the transfer cost defined in \eqref{Eqn: Mod_Centralized}, and the MGs
decomposed subproblems defined in \eqref{Eqn: Subproblem} solved with
the distributed algorithm, are analyzed next.

\begin{table}[tbp]
\caption{MMG system data for the base case}
\begin{center}
\begin{tabular}{lccc}
\hline
\textbf{Parameter}  & \textbf{MG1}  & \textbf{MG2}  & \textbf{MG3}  \\
\hline
$a$ $ [\mathrm{\$/kW^2}] $                  & 0.000132  & 0.0003  & 0.0001  \\
$b$ [\$/kW]                                 & 0.196     & 0.3     & 0.224   \\
$c$ [\$]                                    & 3.548     & 6.105   & 7.5     \\
$D$ [kW]                                    & 210       & 125     & 75      \\
$ \overline{P_{m}^{\textrm{PCC}}} $ [kW]    & 100       & 100     & 100     \\
$ \overline{P}_{m} $ [kW]                   & 150       & 300     & 150     \\
$ c_{d,m} $ [\$/kW]                         & 0.082     & 0.082   & 0.082   \\
$ c_{m,d} $ [\$/kW]                         & 0.05      & 0.05    & 0.05    \\
\hline
\end{tabular}
\end{center}
\label{Tab: Data_Base}
\end{table}

\subsection{Base Case}
\begin{figure}[htbp]
\centering
\includegraphics[scale=1.0]{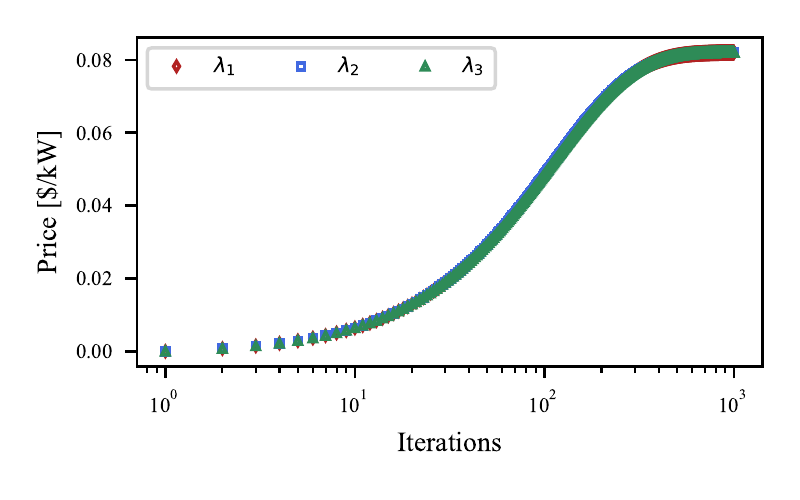}
\vspace{-4ex}
\caption{Convergence of the selling prices for the base case.}
\label{Fig: Selling_Prices}
\end{figure}

\begin{figure}[htbp]
\centering
\includegraphics[scale=1.0]{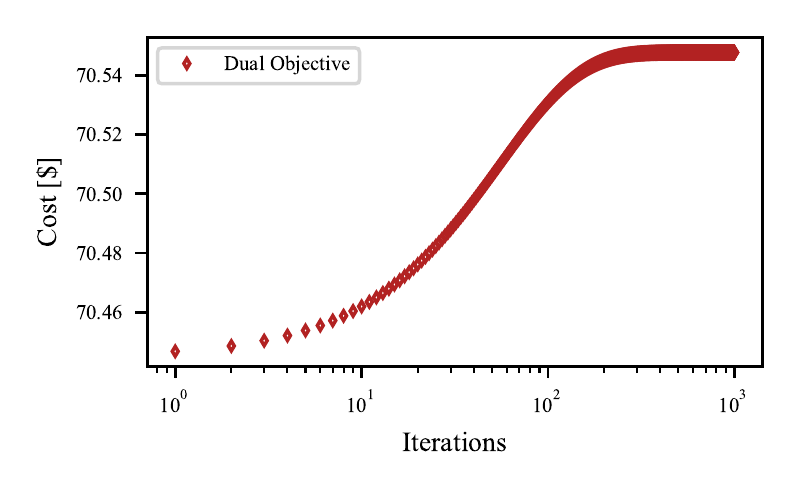}
\vspace{-4ex}
\caption{Convergence of the dual objective for the base case.}
\label{Fig: Dual_Objective}
\end{figure}

\begin{figure}[htbp]
\centering
\includegraphics[scale=1.0]{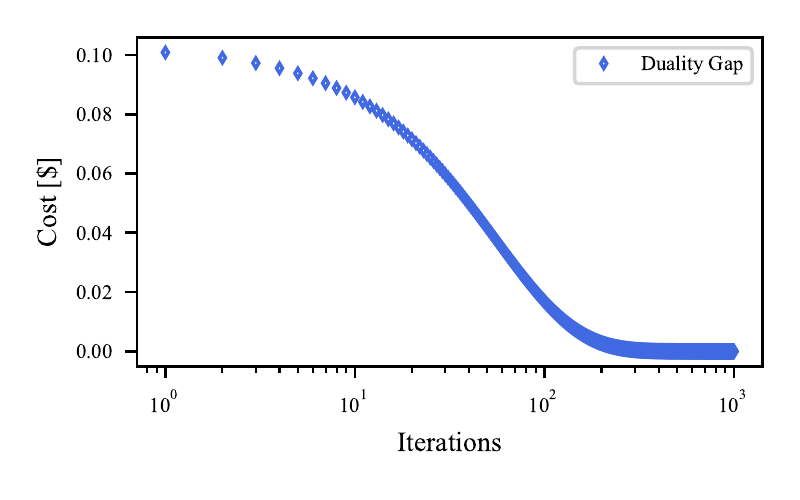}
\vspace{-4ex}
\caption{Convergence of the duality gap for the base case.}
\label{Fig: Duality_Gap}
\end{figure}

Results of the centralized problem, modified centralized problem, and
corresponding distributed solution for the base case are presented in
Table~\ref{Tab: Base_Case}, for an update step $\alpha$ for the subgradient
method of 0.0009 and a scaling factor $\beta$ for the power transfer cost
function of 0.1, chosen through trial and error.
The results show that the distributed algorithm converges to the solution of
the modified centralized problem with sufficient precision.
Figs.~\ref{Fig: Selling_Prices}~to~\ref{Fig: Duality_Gap} show the
convergence of the selling prices, dual objective function, and duality gap
through the iterations, respectively.
Observe in Fig.~\ref{Fig: Duality_Gap} that the duality gap, which is the
difference between the primal and dual objective function values, converges
towards zero, indicating that the distributed algorithm is finding the optimal
solution of the centralized problem.
In this particular case, there are no power exchanges
among MGs in the modified centralized and distributed solutions, as shown
in Table~\ref{Tab: Base_Case}. For this reason, the objective function values
of the centralized, modified centralized, and distributed problems are all
equal.

The iterative algorithm is relatively fast, requiring 29.84 seconds to complete
the predefined 1,000 iterations, which is reasonable, since the scheduling
period is assumed to be one hour. A fixed number of iterations was chosen to
visualize the convergence more clearly. However, using a different convergence
criterion, such as the duality gap expressed in percentage, will require less
iterations, as can be seen in Fig.~\ref{Fig: Duality_Gap}.
The selling prices could also be initialized at different values, if a previous
solution is known, which improves the convergence rate of the algorithm.
However, since these values may not always be available, it is important to
verify the convergence of the algorithm with arbitrary intial values; hence,
the initial values of $\lambda_{m}$ were assumed to be zero.

\begin{table}[htbp]
\caption{Base case results}
\begin{center}
\begin{tabular}{llll}
\hline
& \textbf{Centralized}
& \begin{tabular}[c]{@{}l@{}}\textbf{Modified} \\
        \textbf{Centralized}\end{tabular}
& \textbf{Distributed} \\
\hline
Objective [\$]        & 70.54   & 70.54   & 70.54   \\
Generation Cost [\$]  & 47.99   & 47.99   & 47.99   \\
Transfer Cost [\$]    & --      & 0.00    & 0.00    \\
$P_{1}$               & 110.00  & 110.00  & 110.00  \\
$P_{2}$               & 25.00   & 25.00   & 25.00   \\
$P_{3}$               & 0.00    & 0.00    & 0.00    \\
$P_{1,2}^{e}$         & 0.00    & 0.00    & 0.00    \\
$P_{1,3}^{e}$         & 0.00    & 0.00    & 0.00    \\
$P_{2,1}^{e}$         & 0.00    & 0.00    & 0.00    \\
$P_{2,3}^{e}$         & 0.00    & 0.00    & 0.00    \\
$P_{3,1}^{e}$         & 8.05    & 0.00    & 0.00    \\
$P_{3,2}^{e}$         & 8.05    & 0.00    & 0.00    \\
$P_{d,1}$             & 91.95   & 99.99   & 99.99   \\
$P_{d,2}$             & 91.95   & 99.99   & 99.99   \\
$P_{d,3}$             & 91.10   & 75.00   & 75.00   \\
$P_{1,d}$             & 0.00    & 0.00    & 0.00    \\
$P_{2,d}$             & 0.00    & 0.00    & 0.00    \\
$P_{3,d}$             & 0.00    & 0.00    & 0.00    \\
$\lambda_{1}$         & 0.22    & 0.22    & 0.082   \\
$\lambda_{2}$         & 0.31    & 0.31    & 0.082   \\
$\lambda_{3}$         & 0.082   & 0.082   & 0.082   \\
\hline
\end{tabular}
\end{center}
\label{Tab: Base_Case}
\end{table}

\subsection{Stressed Case}
This case was designed to ensure that power exchanges among MGs occur, to
study the performance of the distributed algorithm under such conditions.
Thus, the following modifications to the data were made: increased the demand
of MG2 from 125~kW to 325~kW; decreased the maximum generation of MG2 from
300~kW to 150~kW; increased the maximum generation of MG1 and MG3 from 150~kW
to 200~kW; made generation of MG2 extremely expensive by increasing its linear
cost coefficient from 0.3 to 30; increased the PCC capacity limit of MG2 from
100~kW to 200~kW (necessary for feasibility); and made the main grid selling
prices extremely high, i.e., changed from 0.082~\$/kW to 82~\$/kW, to ensure
that MG2 buys power from other MGs.

Table~\ref{Tab: Data_Stressed} shows the corresponding data considering the
aforementioned modifications, with $\alpha = 0.009$ and $\beta = 0.1$.
In this case, there are now power exchanges among MGs, as shown in
Table~\ref{Tab: Stressed_Case}; for example, there is a power transfer of
26.67~kW between MG1 and MG2. The power transfer cost in the
objective functions of the modified centralized and distributed problems
represents around 5.2\% of the objective function value of the centralized
problem, and could be regarded as the tradeoff for attaining a decomposable
problem suitable for solution with the distributed algorithm.

\begin{figure}[htbp]
\centering
\includegraphics[scale=1.0]{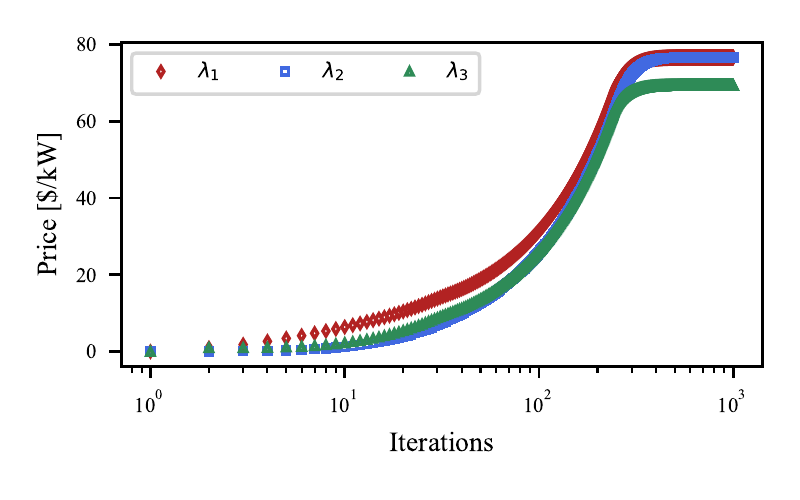}
\vspace{-3ex}
\caption{Convergence of the selling prices for the stressed case.}
\label{Fig: Selling_Prices_Stressed}
\end{figure}

\begin{figure}[htbp]
\centering
\includegraphics[scale=1.0]{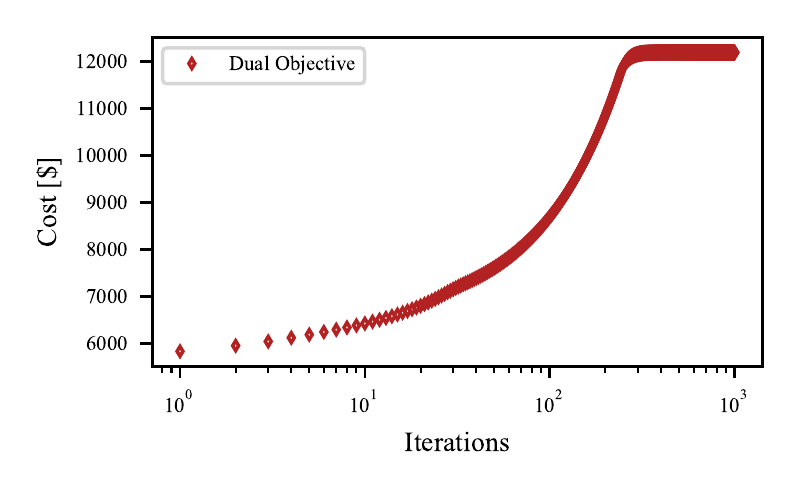}
\vspace{-3ex}
\caption{Convergence of the dual objective for the stressed case.}
\label{Fig: Dual_Objective_Stressed}
\end{figure}

\begin{figure}[htbp]
\centering
\includegraphics[scale=1.0]{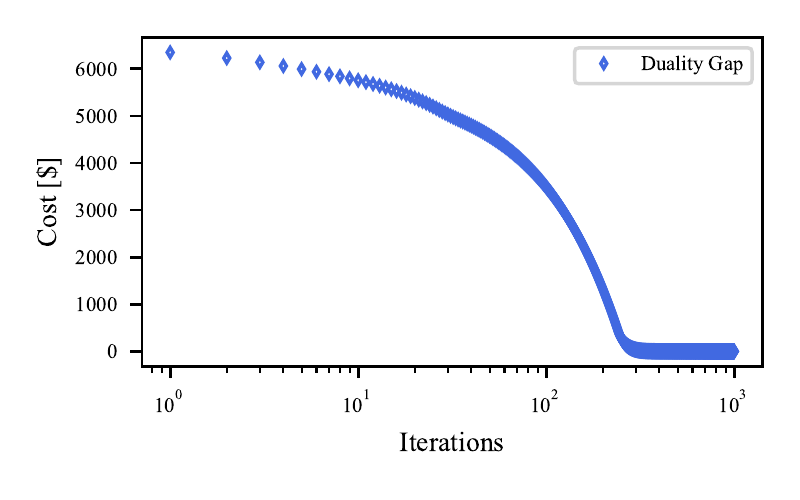}
\vspace{-3ex}
\caption{Convergence of the duality gap for the stressed case.}
\label{Fig: Duality_Gap_Stressed}
\end{figure}

Once again, the distributed algorithm converges towards the solution of the
centralized modified problem. Thus,
Figs.~\ref{Fig: Selling_Prices_Stressed}~to~\ref{Fig: Duality_Gap_Stressed}
show the convergence of the selling prices, dual objective, and duality gap
through the iterations, respectively. The duality gap converges towards zero
as in the previous case, with 1,000 iterations being completed in 23.92
seconds, which is again a reasonable execution time.

\begin{table}[htbp]
\caption{MMG system data for the stressed case}
\begin{center}
\begin{tabular}{lccc}
\hline
\textbf{Parameter}  & \textbf{MG1}  & \textbf{MG2}  & \textbf{MG3}  \\
\hline
$a$ $ [\mathrm{\$/kW^2}] $                & 0.000132  & 0.0003  & 0.0001    \\
$b$ [\$/kW]                               & 0.196     & 30      & 0.224     \\
$c$ [\$]                                  & 3.548     & 6.105   & 7.5       \\
$D$ [kW]                                  & 210       & 325     & 75        \\
$ \overline{P^{\textrm{PCC}}}_{m} $ [kW]  & 100       & 200     & 100       \\
$ \overline{P}_{m} $ [kW]                 & 200       & 150     & 200       \\
$ c_{d,m} $ [\$/kW]                       & 82        & 82      & 82        \\
$ c_{m,d} $ [\$/kW]                       & 0.05      & 0.05    & 0.05      \\
\hline
\end{tabular}
\end{center}
\label{Tab: Data_Stressed}
\end{table}

\begin{table}[htbp]
\caption{Stressed case results}
\begin{center}
\begin{tabular}{llll}
\hline
& \textbf{Centralized}
& \begin{tabular}[c]{@{}l@{}}\textbf{Modified} \\
        \textbf{Centralized}\end{tabular}
& \textbf{Distributed} \\
\hline
Objective [\$]        & 11580.64  & 12187.31  & 12187.31  \\
Generation Cost [\$]  & 4610.64   & 4610.64   & 4610.64   \\
Transfer Cost [\$]    & --        & 606.67    & 606.67    \\
$P_{1}$               & 200.00    & 200.00    & 200.00    \\
$P_{2}$               & 150.00    & 150.00    & 150.00    \\
$P_{3}$               & 175.00    & 175.00    & 175.00    \\
$P_{1,2}^{e}$         & 46.30     & 26.67     & 26.67     \\
$P_{1,3}^{e}$         & 0.00      & 0.00      & 0.00      \\
$P_{2,1}^{e}$         & 0.00      & 0.00      & 0.00      \\
$P_{2,3}^{e}$         & 0.00      & 0.00      & 0.00      \\
$P_{3,1}^{e}$         & 29.36     & 36.67     & 36.67     \\
$P_{3,2}^{e}$         & 70.63     & 63.33     & 63.33     \\
$P_{d,1}$             & 26.94     & 0.00      & 0.00      \\
$P_{d,2}$             & 58.06     & 85.00     & 84.99     \\
$P_{d,3}$             & 0.00      & 0.00      & 0.00      \\
$P_{1,d}$             & 0.00      & 0.00      & 0.00      \\
$P_{2,d}$             & 0.00      & 0.00      & 0.00      \\
$P_{3,d}$             & 0.00      & 0.00      & 0.00      \\
$\lambda_{1}$         & 81.99     & 76.67     & 76.67     \\
$\lambda_{2}$         & 82.00     & 82.00     & 76.67     \\
$\lambda_{3}$         & 0.259     & 0.259     & 69.33     \\
\hline
\end{tabular}
\end{center}
\label{Tab: Stressed_Case}
\end{table}

\subsection{Individual and Cooperative Operation}
A final experiment was carried out to compare the costs of MGs operating on
their own, with no power exchanges, against the cost when there are power
exchanges. For this test, the stressed case data was used, with the modification
of increasing the maximum generation capacity of MG1 and MG2, from 200~kW and
150~kW to 250~kW and 350~kW, respectively, to ensure feasibility when there are
no power exchanges, and the PCC limits of all MGs were set to zero to restrict
the power exchanges.
The results are shown in Table~\ref{Tab: Cost_Comp_MGs_MMGs}, where the first
column shows the operation costs of each MG and the complete system, when
operating individually, and the second column shows the operation costs when
power exchanges are enabled. The third column shows the difference between
the individual and cooperative operation modes, which represents the
improvement in costs achieved by exchanging power.
Note that all MGs improve their individual operation costs as a consequence of
the exchanges, which reduces the total cost of the system from \$9,863.17 to
\$6,716.93 (around 32\%).
The degree of improvement will depend on the generation costs of each MG,
their demand, the costs of buying and selling electricity from and to the
main grid, and the PCC limits. The possibility of power exchanges allows
MGs to benefit from higher or lower costs to meet their demand.

\begin{table}[htbp]
\caption{Individual MGs vs. Cooperative MGs}
\begin{center}
\begin{tabular}{lccc}
\hline
& \textbf{Individual}  & \textbf{Cooperative}  & \textbf{Difference}  \\
\hline
MG1 [\$]         & 50.52     & -980.68   & 1,031.20  \\
MG2 [\$]         & 9,787.79  & 7,673.75  & 2,114.04  \\
MG3 [\$]         & 24.86     & 23.86     & 1.00      \\
Total Cost [\$]  & 9,863.17  & 6,716.93  & 3,146.24  \\
\hline
\end{tabular}
\end{center}
\label{Tab: Cost_Comp_MGs_MMGs}
\end{table}

\section{Conclusion}

This paper presented a model for the optimal operation of an MMG system that
considers power exchanges among a set of MGs and the main grid. It was assumed
that the interaction among MGs occur through the main grid, and therefore the
MGs PCC could limit the amount of power exchange.
The model was decomposed using Lagrangian relaxation and solved through an
iterative distributed procedure using the subgradient method to find the
solution of the corresponding dual problem, thus preserving the generation cost
and demand information of each MG.
Results showed that the distributed algorithm converges to the optimal solution
with high precision within a reasonable execution time, making the proposed
model a viable alternative for implementing a distributed EMS in an MMG system.
Furthermore, the operation costs of all MGs improved when engaging in power
exchanges, as opposed to their individual operation.

The developed formulation was based on a simplified model, designed to focus
on the decomposition method and the benefits of the distributed approach,
to facilitate their understanding. Future work will include more detailed
modeling to consider relevant components in the MG, as well as uncertainties
in renewable generation and demand.
Furthermore, the convergence performance of the subgradient method could be
improved by applying parallel processing techniques, which may significantly
reduce execution times. Other methods for updating the multipliers could be
also explored.

%%%%% BIBLIOGRAPHY
\bibliographystyle{IEEEtran}
\bibliography{references}

\end{document}